\documentclass[showpacs,twocolumn,prl]{revtex4}
\usepackage{amsmath}
\usepackage{amssymb}
\usepackage{color}
\usepackage{graphics}
\usepackage{epsfig}
\usepackage{floatflt}
\usepackage{hyperref}
\usepackage{epsfig}

\begin{document}

\title {Anomalous temperature dependence of the Casimir force for thin metal films}

\author{V.~A.~Yampol'skii$^{1,2}$, Sergey Savel'ev$^{1,3}$, Z. A. Mayselis$^{2}$, S. S. Apostolov$^{2}$, Franco Nori$^{1,4}$}
\affiliation{$^1$Frontier Research System, The Institute of
Physical and Chemical Research (RIKEN), Wako-shi, Saitama,
351-0198, Japan
\\ $^2$A. Ya. Usikov Institute for Radiophysics and
Electronics National Academy of Sciences of Ukraine, 61085
Kharkov, Ukraine \\ $^3$Department of Physics, Loughborough
University, Loughborough LE11 3TU, UK \\
$^4$Department of Physics, Center for Theoretical Physics, Applied
Physics Program, Center for the Study of Complex Systems,
University of Michigan, Ann Arbor, MI 48109-1040, USA}

\date{\today}
\begin{abstract}
{Within the framework of the Drude dispersive model, we predict an
unusual non-monotonous temperature dependence of the Casimir force
for thin metal films. For certain conditions, this force decreases
with temperature due to the decrease of the metallic conductivity,
whereas the force increases at high temperatures due to the
increase of the thermal radiation pressure. We consider the
attraction of a film to: either (i) a bulk ideal metal with a
planar boundary, or (ii) a bulk metal sphere (lens). The
experimental observation of the predicted non-monotonous
temperature dependence of the Casimir force can put an end to the
long-standing discussion on the role of the electron relaxation in
the Casimir effect.}
\end{abstract}  \maketitle

The Casimir effect is one of the most interesting macroscopic
manifestation of the zero-point vacuum oscillations of the quantum
electromagnetic field. This effect manifests itself as the
attractive force arising between two uncharged bodies placed in
the vacuum due to the difference of the zero-point oscillation
spectrum in the absence and in the presence of them (see, e.g.,
the monographs~\cite{m1,m3} and review papers~\cite{r1,r3}).

The Casimir effect attracts considerable attention because of its
numerous applications in quantum field theory, atomic physics,
condensed matter physics, gravitation and
cosmology~\cite{m1,m3,r1,r3,f}. The noticeable progress in the
measurements of the Casimir force~\cite{exp} has opened the way
for various potential applications in nanoscience~\cite{nan},
particularly, in the development of nano-mechanical
systems~\cite{m3,r3,nan}.

In spite of intensive studies on the Casimir effect, it is
surprising that such an important problem as the temperature
dependence of this effect is still unclear and is still an issue
of lively discussion~\cite{a-dr,dr,dr1}. The central point in this
discussion is if the Lifshitz formula (see, e.g., ~\cite{lif}) is
applicable or not \emph{for lossy media}. The authors of
Ref.~\onlinecite{a-dr} have argued that the Drude dispersion
relation for a lossy medium leads to inconsistencies because the
reflection coefficient $r_{\rm TE}$ for the TE electromagnetic
mode becomes discontinuous when the imaginary frequency $\zeta =
-i \omega$ tends to zero. Therefore, instead of the Drude
dispersion relation for the high-frequency dielectric permittivity
$\varepsilon$,
\begin{equation}\label{1}
  \varepsilon(i\zeta)=1+\frac{\omega_p^2}{\zeta (\zeta + \nu)},
\end{equation}
where $\omega_p$ and $\nu$ are the plasma frequency and the
relaxation frequency, authors of Ref.~\onlinecite{a-dr} suggest
the same equation, but with $\nu =0$. Bostr${\rm \ddot{o}}$m and
Sernelius~\cite{dr} have been the first to inquire whether this
prescription is correct. They argued that in view of a realistic
dispersion relation, the TE mode should not contribute to the
Casimir force at zero temperature. Later, the authors of
Refs.~\onlinecite{dr,dr1} have shown that the mentioned
discontinuity of $r_{\rm TE}$ at $\zeta \rightarrow 0$ does not
lead to any physical difficulty or ambiguity.

For the case of zero temperature, the essence of the problem can
be reduced to the following fundamental question: Can the Casimir
force \emph{``feel'' the dissipation parameter} (the relaxation
frequency $\nu$) at zero temperature when \emph{the dissipation
itself is absent}? According to Ref.~\onlinecite{a-dr}, the answer
is ``no''. However, the authors of Refs.~\onlinecite{dr,dr1}
conclude that the answer should be ``yes''.

In this paper, we pay attention to an important feature of the
Casimir force that can be demonstrated within the frame of the
Drude dispersion model. There exist two competing phenomena that
obviously determine the temperature dependence of the Casimir
force. On the one hand, an increase in temperature leads to an
increase of the relaxation frequency and, therefore, to a decrease
of the metal conductivity and to a \emph{ decrease} of the Casimir
force. On the other hand, when increasing the temperature, the
Casimir force \emph{increases } due to the growth of the thermal
radiation pressure. The competition of these two effects can
result in a \emph{non-monotonous} temperature-dependence of the
Casimir force.

The experimental observation of such an anomalous temperature
dependence of the Casimir force might be a direct justification of
the applicability of the Drude model. However, this temperature
effect\emph{ for bulk metals} is very difficult to observe because
of its small magnitude. Indeed, the relative contribution $|F_{\rm
rad}/F(T=0)|$ of the thermal radiation force $F_{\rm rad}$ into
the Casimir force $F$ is proportional to $(kT a/\hbar c )^4 \ll 1$
for realistic distances $a \ll a_T = \hbar c/kT$ between bulk
metals. Here $k$ is the Boltzman constant and $c$ is the speed of
light. The temperature-dependent part of the term $F_\nu$, related
to the relaxation frequency $\nu$, is very small because it is
proportional to the small surface impedance of a metal. Therefore,
for bulk samples, the Casimir force is observed to be \emph{slowly
increasing} with $T$ due to an increase of the radiation term
$F_{\rm rad}$.

In this paper, we predict a \emph{decreasing} Casimir force with
$T$ and show that the difficulties mentioned above, for the
observation of the anomalous temperature dependence of the Casimir
force, can be significantly diminished if we consider the
interaction of \emph{thin metal films}, instead of only between
bulk samples. As was derived in Ref.~\onlinecite{film}, the
temperature effects in the Casimir force can be brought to the
forefront if the film thickness $d$ is the smallest parameter with
the dimension of length. The characteristic frequency $\omega_c$
of the fluctuations, that provide the main contribution to the
Casimir force, becomes smaller,
\begin{equation}\label{2}
\omega_c = \omega_p\sqrt{\frac{d}{a}} \ll \omega_p, \quad \omega_c
\ll \frac{c}{a},
\end{equation}
if
\begin{equation}\label{3}
d \ll \delta = c/\omega_p, \quad \delta \ll a.
\end{equation}
This means that the high-temperature regime for the Casimir
attraction of a film occurs at lower temperatures: at $T > T_c =
\hbar \omega_c/k \propto d^{1/2}$. In addition, under conditions
(\ref{3}), the surface impedance of a metal film is not small.
Therefore, the Casimir force for thin films becomes smaller than
for bulk materials (see results of recent
experiment~\cite{film-exp} with thin films), and the relative role
of the temperature effects in the Casimir force becomes stronger.
Thus, as we demonstrate below, the anomalous temperature
dependence of the Casimir force can be observed, in principle, for
thin metal films. The successful implementation of this experiment
could put an end to the longstanding discussion on the role of the
electron relaxation in the Casimir effect.

{\it Model.---} The general formula for the Casimir interaction
force between dielectric slabs with arbitrary dielectric constants
$\varepsilon$ was originally derived by Lifshitz~\cite{lif1} (see,
also, Refs.~\onlinecite{sh,sh1}). The Casimir force is presented
in this formula as a functional defined on the set of functions
$\varepsilon (i\omega_n)$ of a discrete variable $\omega_n =2\pi
nkT$ ($n=0,1,2, \dots$). For the dielectric permittivity of the
metal film, we choose the Drude dispersive model Eq.~(\ref{1})
which takes into account the temperature dependence of the
relaxation frequency $\nu$ caused by the scattering of electrons
by phonons. We use the relation,
\[
\nu(T)= \nu_0 +\nu_{\rm ph}(T/\Theta),
\]
\vspace{-0.5cm}
\begin{equation}\label{4}
\nu_{\rm ph}(x)=A\,\nu_{\rm ph}(1)\, x^5\int_0^{1/x} \frac{y^5
dy}{\left({\rm e}^y-1\right)\left(1-{\rm e}^{-y}\right)},
\end{equation}
based on the Gr${\rm \ddot{u}}$neisen formula for the temperature
dependence of the resistivity (see, e.g.,
Ref.~\onlinecite{ibach}). Here $\nu_0$ is the residual relaxation
frequency caused by the electron scattering on crystal defects,
$\Theta$ is the Debye temperature, $\nu_{\rm ph}(T/\Theta)$ is the
relaxation frequency due to the electron-phonon scattering. The
value $\nu_{\rm ph}(1)$ depends on the Fermi velocity of
electrons, the strength of the electron-phonon interaction, etc.
This $\nu_{\rm ph}(1)$ can be be obtained by measuring the
resistivity at the Debye temperature. The constant $A$ is $\left(
\int_0^{1} y^5 dy/\left({\rm e}^y-1\right)\left(1-{\rm
e}^{-y}\right)\right)^{-1} \approx 3$. For simplicity, we do not
take into account the surface scattering of electrons in the
explicit form (\ref{4}) because it only changes the value of
$\nu_0$ (see, e.g., Ref.~\onlinecite{mak}).

We consider first the Casimir effect for an ideal bulk conductor
and a thin metal film of thickness $d$, separated by a distance
$a$. Then, using the ``Proximity Force Theorem''~\cite{PrT}, we
derive the expressions for the Casimir force between a metal film
and an ideal metal sphere (lens). The geometry of problem is shown
in Fig.~1.
\begin{figure}[hbpt]
\vspace*{-2cm}
\includegraphics[width=9cm]{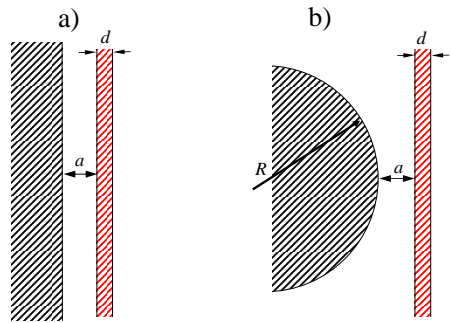}
\vspace*{-7.3cm} \caption{(Color online) Geometry of the problem.
(a) The Casimir attraction of a thin metal film (in red) to an
ideal plane bulk metal. (b) The same film now interacting with a
metal sphere. }\label{f1}
\end{figure}

{\it Casimir force.---} The asymptotic equation for the Casimir
attraction of a \emph{thin} metal film to an ideal \emph{bulk}
plane metal was derived in Ref.~\onlinecite{film}. The force $f$
per unit area can be written in the form,
\begin{equation}\label{2f}
f=-\,\frac{B kT}{8\pi a^{3}}\int_0^\infty d x \, x^3 \mathrm {\rm
e}^ {-x} I(x),
\end{equation}
where $B=(\hbar \,\omega _c/4 \pi kT)^2$,  $\omega_c=\omega_p
(d/a)^{1/2}$,
\begin{equation}\label{3f}
I(x)={\sum_{n=0}^{\infty}} {\vphantom{\sum}} ' \frac{1}{n(n+C)+B
\Psi(x)},
\end{equation}
$\Psi(x)=x(1-\mathrm e^{-x})$,   $C=\hbar\, \nu(T)/2 \pi kT$, the
prime over the sum symbol indicates that the term with $n=0$ is
taken with half the weight. Equation (\ref{2f}) is valid if
conditions (\ref{3}) are fulfilled. In this case, one can neglect
the relativistic retarding effect and pass to the limit $c
\rightarrow \infty$.

Using the Abel-Plana formula for summing series, we can rewrite
Eq.~(\ref{2f}) in the form of a sum of two terms that correspond
to two sources for the temperature dependence of the Casimir
force,
\begin{equation}\label{3bf}
f=f_\nu+f_{\rm rad},
\end{equation}
\vspace{-0.5cm}
\begin{equation}\label{4f}
f_\nu=-\,\frac{\hbar \omega_c}{32\pi^2 a^3} \int_0^\infty d
x\,x^3\mathrm e^{-x} \int_0^\infty \frac{d
\tau}{\tau(\tau+\eta)+\Psi(x)},
\end{equation}
\vspace{-0.5cm}
\[
f_{\rm rad}=-\frac{\hbar\,\nu(T)}{8\pi^2 a^3}\int_0^\infty d x\,
x^3 \mathrm e^{-x}
\]
\vspace{-0.5cm}
\begin{equation}\label{5f}
\times \int_0^\infty  \frac{t d t}{(\mathrm e^{\gamma
t}-1)\{[t^2-\Psi(x)]^2+\eta^2 t^2\}},
\end{equation}
where $\eta = 2\nu(T)/\omega_c$, $\gamma=\hbar\,\omega_c/2 kT$.

The first term in Eq.~(\ref{3bf}) is provided by the quantum
fluctuations of the electromagnetic field. It depends on
temperature only via $\nu (T)$. Obviously, the modulus of this
term \emph{decreases} when increasing the temperature. The
asymptotics of $f_\nu$ for low and high values of the parameter
$\eta$ are:
\begin{equation}\label{6f}
f_\nu=-\,\frac{i_1 \hbar\,\omega_c }{64 \pi
a^3}\left(1-i_2\frac{\nu (T)}{\omega_c}\right),\qquad\nu\ll\omega
_c,
\end{equation}
\vspace{-0.5cm}
\begin{equation}\label{7f}
f_\nu=-\,\frac{3\hbar\,\omega _c ^2}{16\pi^2 a^3 \nu (T)}
\left[\ln\left(\frac{2\nu(T)}{\omega_c}\right)-i_3\right],\,\, \nu
\gg \omega _c,
\end{equation}
where $i_1 \approx 3.51214$, $i_2=4\zeta (3)/\pi i_1 \approx
0.43578$, $i_3 \approx 0.59272$, and $\zeta (x)$ is the zeta
function.

The second term in Eq.~(\ref{3bf}) is caused by the thermal
fluctuations of the electromagnetic field. Its modulus
\emph{increases} when increasing the temperature. This term has
different asymptotics in different temperature intervals:
\begin{equation}\label{10f}
f_{\rm rad}^{\rm low-T}=-\,\frac{\nu (T)(kT)^2}{24 a^3
\hbar\,\omega_c^2}\ln\left(\frac{\hbar\,\omega_c}{kT}\right), \,
kT\ll \hbar\,\nu, \, \hbar\,\omega_c^2/\nu,
\end{equation}
at low temperatures and
\begin{equation}\label{8f}
f_{\rm rad}^{\rm high-T}=-\,\frac{\zeta (3)}{8\pi}\frac{kT}{a^3},
\, \,kT\gg {\rm min}\left(\hbar\,\omega_c, \,\,
\hbar\,\omega_c^2/\nu \right)
\end{equation}
at high temperatures. In the case $\nu \ll \omega_c$, there exist
the intermediate asymptotics,
\begin{equation}\label{9ff}
f_{\rm rad}^{\rm
intermed-T}=-\,\frac{\zeta(3)}{2\pi}\frac{(kT)^3}{a^3 (\hbar \,
\omega_c) ^2}, \, \,\hbar\,\nu\ll kT \ll\hbar\,\omega_c.
\end{equation}

Using Eqs.~(\ref{6f})--(\ref{9ff}) and the Proximity Force
Theorem~\cite{PrT}, one can easily derive the analogue asymptotics
for the Casimir force $F$,
\begin{equation}\label{Fsp}
F = 2\pi R \int_a^\infty d a' f(a'),
\end{equation}
between an ideal metallic sphere of radius $R$ and a thin metal
film:
\begin{equation}\label{6ff}
F_\nu (\nu\ll\omega _c)=-\,\frac{i_1 \hbar\,\omega_c R}{80
a^2}\left(1-\frac{5i_2}{4}\frac{\nu (T)}{\omega_c}\right),
\end{equation}
\begin{equation}\label{7ff}
F_\nu (\nu \gg \omega _c)=-\,\frac{\hbar\,\omega _c ^2 R}{8\pi a^2
\nu (T)} \left[\ln\left(\frac{2\nu}{\omega_c}\right)-i_3
+\frac{1}{6}\right]\,.
\end{equation}
For the low-temperature interval in Eq.~(\ref{10f}),
\begin{equation}\label{10ff}
F_{\rm rad}^{\rm low-T}=-\,\frac{\pi R\nu (T)(kT)^2}{12 a^2
\hbar\,\omega_c^2}\left[\ln\left(\frac{\hbar\,\omega_c}{kT}\right)-\frac{1}{2}\right],
\end{equation}
for the high-temperature interval in Eq.~(\ref{8f}),
\begin{equation}\label{8ff}
F_{\rm rad}^{\rm high-T}=-\,\frac{\zeta (3)}{8}\frac{R kT}{a^2},
\end{equation}
and for the intermediate temperature interval in Eq.~(\ref{9ff}),
\begin{equation}\label{9f}
F_{\rm rad}^{\rm intermed-T}=-\,\zeta(3)\frac{R(kT)^3}{a^2 (\hbar
\, \omega_c) ^2}.
\end{equation}

The above results show that the contribution to the Casimir force
from quantum fluctuations always \emph{decreases} when increasing
the temperature. At the same time, the term related to the thermal
radiation always \emph{increases} with temperature. As a result of
this competition, an interesting \emph{non-monotonous} temperature
dependence of the total Casimir force can be observed. Figure 2
shows the dependence of the Casimir force between a metal film
and: (a) an ideal metal semi-space, and (b) a metal sphere (lens),
on temperature, calculated for different values of the parameter
$\nu_{\rm ph}(1)$. The Casimir force is normalized to the force
$F_{\rm bulk}$ between ideal bulk metals for the same separation
$a$. The force $F_{\rm bulk}$ is, $F_{\rm bulk}^{\rm plane} =
\pi^2 \hbar c\, S/240 a^4$, for bulk samples with planar
boundaries of the area $S$, and $F_{\rm bulk}^{\rm sphere} = \pi^3
\hbar c R/360 a^3$ for the attraction between a metal semi-space
and a metal sphere with radius $R$. Note that we use the relations
for $F_{\rm bulk}$ in the \emph{retardation} limit because
$c/\omega_p a \leq 1$. For the same separations $a$, the Casimir
force for a thin film is calculated for the \emph{non-retardation}
regime, since $c/\omega_c a  = c/\omega_p (da)^{1/2} \gg 1$.
\begin{figure}
\includegraphics[width=9cm]{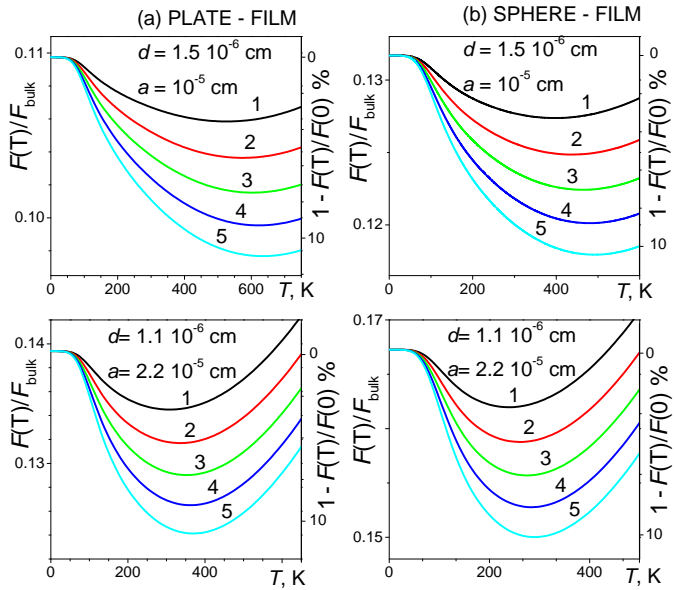} \vspace{-4.2cm}
\caption{The temperature dependence of the Casimir force of
attraction of a metal film to an ideal plane bulk metal (a) and to
a metal sphere (b) for different separations $a$. The left
vertical axes show $F(T)/F_{\rm bulk}$ and the right vertical axes
show $1-F(T)/F(0)$. Curves 1, 2, 3, 4, 5 correspond to $\nu_{\rm
ph}\cdot (10^{-13}$~s) $=0.7,\, 1.05, \,1.4, \,1.75, \,2.1$,
respectively. The values of the other parameters are:
$\omega_p=2\cdot 10^{15}$~s$^{-1}$, $\nu_0=10^{11}$~s$^{-1}$,
$\Theta =100$~K.}
\end{figure}


The decreasing portion of the $F(T)$ dependence corresponds to a
decrease of the Casimir contribution to the entropy when
increasing $T$. This decrease is connected to the enhancement of
the electron scattering on phonons and is much weaker than the
increase of the phonon contribution to the entropy. Thus, the
total entropy certainly increases when increasing the temperature.

In conclusion, we predict an unusual non-monotonous temperature
dependence of the Casimir attraction force between a thin metal
film and a bulk plane ideal metal or a metal sphere (lens).
Usually, for bulk samples, the Casimir force \emph{increases}
slowly with temperature. Here we predict a \emph{noticeable
decrease} of the force with an increase of $T$ for metal films.
The experimental observation of this unusual temperature
dependence of the Casimir force can put an end to the
long-standing dispute on the role of the electron relaxation in
the Casimir effect.

We acknowledge partial support from the NSA, LPS, ARO, NSF grant
No. EIA-0130383, JSPS-RFBR 06-02-91200, MEXT Grant-in-Aid No.
18740224, the EPSRC via No. EP/D072581/1, EP/F005482/1, ESF AQDJJ
network programme, and the JSPS CTC Program.

\end{document}